\documentclass[12pt, A4]{article}
\usepackage[utf8]{inputenc} 
\usepackage
[
        a4paper,
        left=4cm,
        right=2cm,
        top=2cm,
        bottom=2cm        
]
{geometry}
\usepackage{setspace}
\usepackage{subfig} 
\usepackage{graphicx}
\usepackage{amssymb,amsmath}
\usepackage{amsmath}
\usepackage[export]{adjustbox}
\usepackage{float}
\usepackage{gensymb}\usepackage{tikz}
\usepackage{multirow}
\usepackage{longtable}
\usepackage{listing}
\usepackage{color}

\providecommand{\pacs}[1]{\textbf{\textit{PACS number:}} #1}
\providecommand{\keywords}[1]{\textbf{\textit{keywords:}} #1}

\title{Government intervention modeling in microeconomic company market evolution}
\author{Micha{\l} Chorowski\thanks{ma.chorowski@student.uw.edu.pl}~\thanks{Presented at the 10th Polish Symposium on Physics in Economy and Social Sciences (FENS) 2019, NCBJ Otwock-Swierk.} \thanks{College of Inter Faculty Individual Studies in Mathematics
and Natural Sciences,
University of Warsaw,
Stefana Banacha 2C, PL-02093, Warsaw, Poland} \footnotemark[4]~~and Ryszard Kutner\thanks{Faculty of Physics, University of Warsaw, Pasteur Str. 5, PL-02093 Warsaw, Poland} }

\date{}
 
\begin{document}
 
\begin{titlepage}

\maketitle

\begin{abstract}
{Modern technology and innovations are becoming more crucial than ever for the survival of companies in the market. Therefore, it is significant both from theoretical and practical points of view to understand how governments can influence technology growth and innovation diffusion (TGID) processes. We propose a simple but essential extension of Ausloos-Clippe-P\c{e}kalski and related Cichy numerical models of the TGID in the market. Both models are inspired by the nonlinear non-equilibrium statistical physics. Our extension involves a parameter describing the probability of government intervention in the TGID process in the company market. We show, using Monte Carlo simulations, the effects interventionism can have on the companies' market, depending on the segment of firms that are supported. The high intervention probability can result, paradoxically, in the destabilization of the market development. It lowers the market's technology level in the long-time limit compared to markets with a lower intervention parameter. We found that the intervention in the technologically weak and strong segments of the company market does not substantially influence the market dynamics, compared to the intervention helping the middle-level companies. However, this is still a simple model which can be extended further and made more realistic by including other factors. Namely, the cost and risk of innovation or limited government resources and capabilities to support companies.} 
\end{abstract}

\\


\end{titlepage}
 
\addtocounter{page}{1}
\section{Introduction}

New technologies have a tremendous impact on our economy. Ongoing globalization opened a new space for competition between firms all over the world, and it fuels the race for the most innovative technologies. Nowadays, the most technologically advanced firm is often the winner. The new technologies change how particular firms operate and perform, so it is interesting to study how technology can influence the market as a whole. This impact can not be underestimated, and it is no surprise that physicists are also interested in this many-body problem. A non-linear statistical physics simulated model for technological growth was recently proposed by Ausloos, Clippe, and P\c{e}kalski (ACP model) \cite{ausloos2004model} and next extended by Cichy \cite{cichy} (Ci-model).

Our work is directly inspired by Ci-model. It describes the influence of technology diffusion on market growth dynamics. The central assumption is that the companies' survival depends only on their technology level. It might be a simplistic assumption; nonetheless, the results agree with the empirical data \cite{cichy}. Therefore, we also make this assumption. However, we expand the model to make it more general, versatile, and realistic by taking into account the interventionism.

The companies operate in certain countries, and these countries' politics influence the way those firms operate. The government and its policy can change the firms' behavior, and those changes should be visible on the aggregate level as well. We propose a model that expands the existing technological growth model \cite{cichy} to allow for modeling the influence of government intervention on company market growth. We do so by using a single parameter $0\leq q\leq 1$, which is the probability of the government intervention in case of a firm's incoming bankruptcy. In practice, the intervention can be realized by government stimuli such as, for example, subsidies, grants, tax reliefs, taking over the company's majority stake, and ultimately even by nationalizing the company. We also consider how the influence of this intervention differs when the government targets only specific segments of the company market.

It should be emphasized that the problem of state interventionism in the free market of companies is still at an early stage, although this problem is as old as the free market \cite{aikins2009political,datta1990market,napoles2014macro}.

\section{Method}

\subsection{Model}

We define the model as follows. In the beginning, a two-dimensional rectangular lattice of size $L_x \cdot L_y$ is created ($L_x=L_y=10$ in our case for simplicity). It is then occupied with pairs of numbers that represent the states of firms in the market. Two numbers describe each firm (e.g., the $i$th): (i) its market share $\omega_i\geq 0$ and (ii) its technology level $A_i\geq 0$ -- the latter is initially a random number between 0 and 1 drawn from a uniform distribution. The initial lattice density was chosen to be $0< c\leq 1$, which means that $c$ percent of the lattice sites are occupied with firms at the beginning of the simulation. Other lattice sites are empty. For example, we chose  $c=0.8$ (the same as in \cite{cichy}) as a representative for the higher concentration case. We plan to take into account other concentrations in   future work. We decided to consider the higher concentration case ($c>0.5$) as it is more promising than the dilute case ($c\leq 0.5$) in respect to a possible dynamical phase transition.

The market shares of all the companies have to sum up to 1 at any moment according to the normalization condition,
\begin{equation}
    \sum_j \omega_j (t) = 1, ~~\textrm{for any time}~ t.
   \label{eq:norm}
   \end{equation}
 Initially all of the companies have equal shares $\omega_i (0) = 1/N(0)$, where $N(t)$ is the number of firms at a given time (Monte Carlo step/site defined further in the text) $t$.

In every time-step, the system is characterized by some average technology level weighted by companies' shares,
\begin{equation}
\langle A(t)\rangle = \sum_{j=1}^{N(t)} \omega_j A_j(t),
\label{eq:A}
\end{equation}
with the initial condition, e.g. $\langle A(0) \rangle \approx 0.5$. This seems to be optimal in relation to the initial condition of world frontier dynamics given by Eq. (\ref{eq:F}) below.

As in paper \cite{cichy}, the companies in the market can benefit from copying the world frontier whose technology is described by 
\begin{equation}
F(t) = e^{\sigma t},
\label{eq:F}
\end{equation}
where $\sigma $ is the parameter measuring the world technological progress rate. In this model, the growth of the world's technological advancement is defined by Moore's law of exponential growth \cite{schaller1997moore}. In our model we assume (as the most general case) that initially $\langle A(t=0)\rangle < F(t=0)$. We found a particularly interesting phase appearing at the transition time $t_c$, which we define as the first time-step for which the inequality $\langle A(t)\rangle > F(t=0)$ is fulfilled. We find $t_c$ numerically by calculating $\langle A(t)\rangle$ averaged additionally over time-dependent statistical ensemble of simulations (or replics of system). We show how the value of $t_c$ depends on the probability of government intervention $q$. The changes in the market evolution that come right after $t_c$ are an essential aspect of our model that has both theoretical and practical meaning.

We note that the model parameters such as the current number of companies operating on the market      $(N(t))$, the shares of individual companies in the market $(\omega _j)~,~j=1,2,\ldots ~,N(t)$,
the world technological frontier progress rate (e.g., of the European Union) $(\sigma )$, and even the likelihood of mergers of companies $(b)$ can be obtained from empirical data. The analogue of inverse market temperature $(s)$, (auxilliary) concentration $(c)$, and the minimal number of companies needed for the market to work $(N_{min})$ are free parameters of the model that can be obtained by matching the relative level of technology  $(\langle A(t)\rangle /F(t))$ with empirical data. Let us add that the average level of market technology $\langle A(t)\rangle $, as well as the current technology level of the world leader $(F(t))$, can be measured separately (although it is complicated because it depends on the definition of technology level) \cite{cichy2009human}. As one can see, our model is prepared both for the analysis of empirical data and  extension and generalization -- it is a fully open model.

\subsection{Algorithm}\label{section:algor}

In this paper, we extend the Ci-model \cite{cichy} substantially. This extension is important because it takes into account government interventionism supporting failing companies. This interventionism is described in the model by the probability $0\leq q\leq 1$, which we simply refer to as the parameter of government interventionism. A detailed description of the extended Ci-algorithm is presented below. 

As usual, in the Monte Carlo simulations, we consider two time scales. The first is the Monte Carlo step (MCS or $t_ {MCS}$) in which one cycle of the algorithm is carried out for a selected, single company. The second scale is a Monte Carlo step per lattice site (MCS/site or time step $t$). We define it as such a number of MCSs, which is equal to the number of sites of the substrate lattice.

Our algorithm is as follows:
   \begin{enumerate}
   \item  Randomly pick a single company from $N(t)$ already existing firms by using a uniform distribution. Let it have index $i$.
   \item Calculate the probability of the firm's survival $p_i(t_{MCS})$ (defined by Eq. (\ref{eq:p})).
   \item Compare value $p_i(t_{MCS})$ with a random number from a uniform distribution $r_1 \in (0,1)$. If $r_1 > p_i(t_{MCS})$ the firm bankrupts and disappears from the system leaving its site empty. The shares of the picked firm are equally distributed among all the other firms in the market in such a way that the normalization condition (\ref{eq:norm}) holds. Compared to the original model we modified this step by including government intervention, modeled by the intervention parameter $q$. In this Monte Carlo step, when a firm should bankrupt because $r_1 > p_i(t_{MCS})$, another random number $q_{rnd} \in (0,1)$ is generated and compared with the parameter $q$. If $q_{rnd} \leq q$ the firm is saved by the government and survives at this step despite the fact that $r_1 > p_i(t_{MCS})$. The algorithm goes straight to item 7. If $q_{rnd} > q$ the firm bankrupts anyway. Thus, the parameter $q$ describes the probability that the government will intervene and save the firm (with probability $(1-p_i(t_{MCS}))q$) at risk from bankruptcy. \label{alg:3}
   \item If $r_1 \leq p_i(t_{MCS})$ the firm survives and gets an opportunity to be active. It can move to one of the four randomly chosen neighboring sites. If the chosen site is unoccupied then it does not move. However, it interacts with the firm in that chosen site, in a way described in item 6 below.
   \item If there was an empty site and the firm moved there, the algorithm checks if it has any neighboring firms in the nearest and next-nearest lattice site of its new location. If there are no such firms, then the firm's technology grows according to formula $A_i (t_{MCS}+1) = A_i (t_{MCS}) + r_2 (F(t_{MCS}) - A_i (t_{MCS}))$, where $r_2 \in (0,1)$ is a random number. This growth corresponds to the external technology diffusion since the company copies, although imperfectly, the technology of the world frontier. Of course, if the selected neighboring place was already occupied, the company remains in its site.
   \item There are two mechanisms of inner technology diffusion, merging and creating new firms -- spin-offs. If the $i$-th company found, after moving to a new lattice site, some randomly chosen neighboring firm $j$, then with probability $b$, the firm $i$ merges with the firm $j$. The firm $j$ disappears from the system and the technology of the $i$-th firm changes according to relation $A_i (t_{MCS}+1) = \textrm{max}( A_i (t_{MCS}), A_j (t_{MCS}))$. The $i$-th firm also acquires the shares of the merged firm, so $\omega_i = \omega_i + \omega_j$. With probability $1-b$, firms $i$ and $j$ create a spin-off. In this case, none of the firms disappear from the system. A position $k$ for the spin-off is chosen randomly out of the eight neighboring sites of the $i$-th firm. The spin-off appears, only if the picked site $k$ is unoccupied. The spin-off's technology level is $A_k (t_{MCS}+1) = max( A_i (t_{MCS}), A_j (t_{MCS}) )$ and the shares are $\omega_k = \omega_s (\omega_i + \omega_j)$, where $\omega_s \in (0,1)$. Because the condition \ref{eq:norm} has to be fulfilled, the shares of the company $i$ decrease by product $\omega_i \omega_s$ and the company $j$ by $\omega_j \omega_s$. This step corresponds to the technology diffusion within the companies' market.
   \item The algorithm returns to point 1. until all the $N(t)$ firms have been chosen which ends this Monte Carlo step/site. Next, the algoritm goes to the subsequent Monte Carlo step/site.
  
   \end{enumerate}
   
In the case of a company's bankruptcy, as described in item 3, the shares of the company are distributed equally between other firms in the system. We chose the same way of shares distribution as used in previous model \cite{cichy}. This way of distribution allows for a straightforward comparison of our results with the corresponding in the previous work. It illustrates the role of the intervention parameter, which we introduce into the model. We also assume that after a company bankrupts, the other companies do not perform any additional actions. Their increase in shares is a result of the disappearance of one of the firms from the market and not a result of competing for the shares. Other ways of shares redistribution are also possible, for example, ones in which the freed shares are obtained by other companies proportionally to their present shares (preferential selection rule).

It is also worth realizing that random numbers $r_1$ and $r_2$ are parameterizing the random risk component of investing in innovation by a company.
    
Probability of a company's survival is calculated according to the formula,
  \begin{align} \label{eq:p}
p_i(t_{MCS}) = \left\{ \begin{array}{ll} 
                e^{-sG} & \hspace{-1mm}\textrm{if}~G>0,~\langle A(t_{MCS}) \rangle < 1 \\
                 1 & \hspace{-1mm}\textrm{if}~G<0,~\langle A(t_{MCS})\rangle < 1 \\
                e^{-sH} & \hspace{-1mm}\textrm{if}~H>0,~\langle A(t_{MCS}) \rangle \geq 1 \\
                1 & \hspace{-1mm}\textrm{if}~H<0,~\langle A(t_{MCS})\rangle \geq 1,
                \end{array}\right.
\end{align}
where $G=\langle A(t_{MCS}) \rangle F(t) - A_i(t_{MCS})$ and $H=F(t) - A_i (t_{MCS})$.

Eq. (\ref{eq:p}) divides the technological development of a country (or companies' market) into two phases. The first phase for $\langle A(t) \rangle  < F(t=0)=1$ and the second phase for $\langle A(t) \rangle \geq F(t=0)=1$. The first phase corresponds to the less technologically advanced countries (or companies' markets). The second phase corresponds to countries (or companies' markets) that are more developed and have a higher technology level. The $s$ parameter controls the market vulnerability to technological backwardness and thus susceptibility to bankruptcy. If this parameter has a smaller value, then this vulnerability is smaller and vice versa.  
   
We chose, for example, the same parameters for our simulations as used in paper \cite{cichy}. Those are:
$\sigma = 0.01,~s=1,~b=0.01,~N_{min} = 10,~\omega _s = 0.1$.  
The minimum number of companies $N_ {min}$ necessary for the market survival (i.e., ensure the duration of the simulation) should be calibrated to empirical data. In principle, this choice is optional and the obvious condition $N_ {min}\leq L_x\cdot L_y = 100$ must be met. When the number of firms in the system  equals $N_{min}$ at time $t_{MCS}$, then a randomly picked firm $i$ does not face the danger of bankruptcy. It means that items 2 -- 3 of the algorithm are omitted and the firm goes directly from item 1 to 4.

In our program, the shares of every company are written in a double-precision floating-point format (double). However, it is not an exact representation. To ensure the normalization of market shares with satisfactory accuracy, we additionally introduce a dynamic correction of these shares in each Monte Carlo step per site. It ensures that the normalization condition is met with controlled accuracy (in our case, not exceeding 1\%).

\section{Results and discussion}

\subsection{Free companies' market}

\begin{figure}[H]
\centering
\advance\leftskip-0cm
\includegraphics[width=153mm, height=50mm]{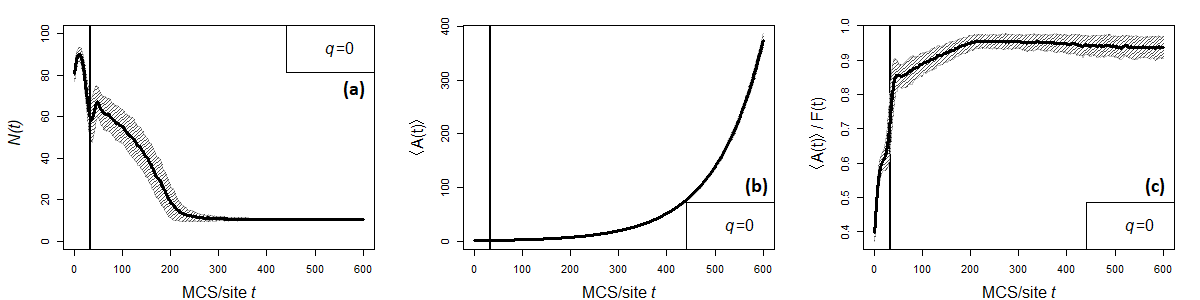}
\caption{Results for the no intervention case (the intervention probability $q=0$). The three plots show the values of the current number of firms $N(t)$ (panel (a)), the average technology $\langle A(t) \rangle$ (panel (b)) and the average to frontier technology ratio $\langle A(t)\rangle /F(t)$ (panel (c)) as functions of time $t$. The vertical straight line marks the time $t_c$ at which $\langle A(t)\rangle \geq F(t=0)=1$ for the first time. The average values over 400 simulations (or replics of the system) are shown with the black curve and the $\pm 1~SD$ over the simulations is marked as the grey band.}
\label{rys:baza}
\end{figure}

Figure \ref{rys:baza} shows the results for the case of no government intervention ($q = 0$) that is, for the free companies market. Three basic quantities are measured at the beginning of every Monte Carlo step per site $t$: (i) the current number of firms in the system $N(t)$, (ii) the weighted average technology of existing firms $\langle A(t) \rangle$, and (iii) the average to frontier technology ratio $\langle A(t)\rangle /F(t)$. The results shown in Fig. 1 are the reference situation for other cases where interventionism is present (i.e., for $q>0$).

In this model, the evolution of the companies' market goes through three stages \cite{cichy}. For the case of free companies' market, the firms with the lowest technology level are eliminated in about 50 time steps (the first stage), and the average technology increases as a result. In the second stage, which lasts for about another 150 steps, the technological advancement of the market comes mostly from internal technology diffusion, namely from spin-offs and merges. In the last (third) stage (that lasts until the end of the simulation), the technology grows exponentially, benefiting from external diffusion -- getting the technology from the world frontier. Apparently, in the third plot of Figure 1, $\langle A(t)\rangle /F (t)$ value is decaying slowly to a stable level after reaching its maximum around MCS/site t = 220.  The relative average $\langle A(t)\rangle /F (t)$ reaches a plateau when the number of firms $N(t)$ reaches its plateau. Both plateaus are determined by parameter $N_{min}$, which we set, for example, equal to 10. In this third stage, when the number of remaining firms is already small, item 5 of the algorithm is the primary source of technology growth. Therefore, as $N(t)$ decays, so does the $\langle A(t)\rangle /F(t)$, since the pool of firms that can copy the leader shrinks. Indeed, all of those stages are well seen in Fig. \ref{rys:baza}.

In Secs. \ref{section:egalit} -- \ref{section:medium} we investigate four significant cases of interventionism. The egalitarian in which all the companies have access to the government's aid, and three other cases in which only the companies with certain technology level can receive support. In the egalitarian case, any company at risk of failure can be chosen to receive another chance thanks to the government intervention and possibly be saved as described in point \ref{alg:3} of the algorithm. In the other three non-egalitarian cases, the government chooses to support only companies with low, medium, or high technology levels. A given company is qualified for one of these three cases when its technology is below one standard deviation $\sigma_g$ from the mean technology in the system, in between $\pm 1$ $\sigma_g$ or above one $\sigma_g$, respectively. We define $\sigma_g$ as:
\begin{equation}
\sigma_g(t) = \sqrt{ \frac{\sum_{i=1}^{N(t)} (A_i(t) - \langle A(t) \rangle)^2}{N(t)} }.
\end{equation}
Several values of $q$ for all of the cases were examined. However, only results for a few characteristic $q$ values are selected in this work to portray changes in the dynamics as $q$ increases within the wide range $0.3\leq q\leq 0.99$.

\subsection{Egalitarian intervention}\label{section:egalit}

\begin{figure}
\centering
\advance\leftskip-0cm
\includegraphics[width=153mm, height=130mm]{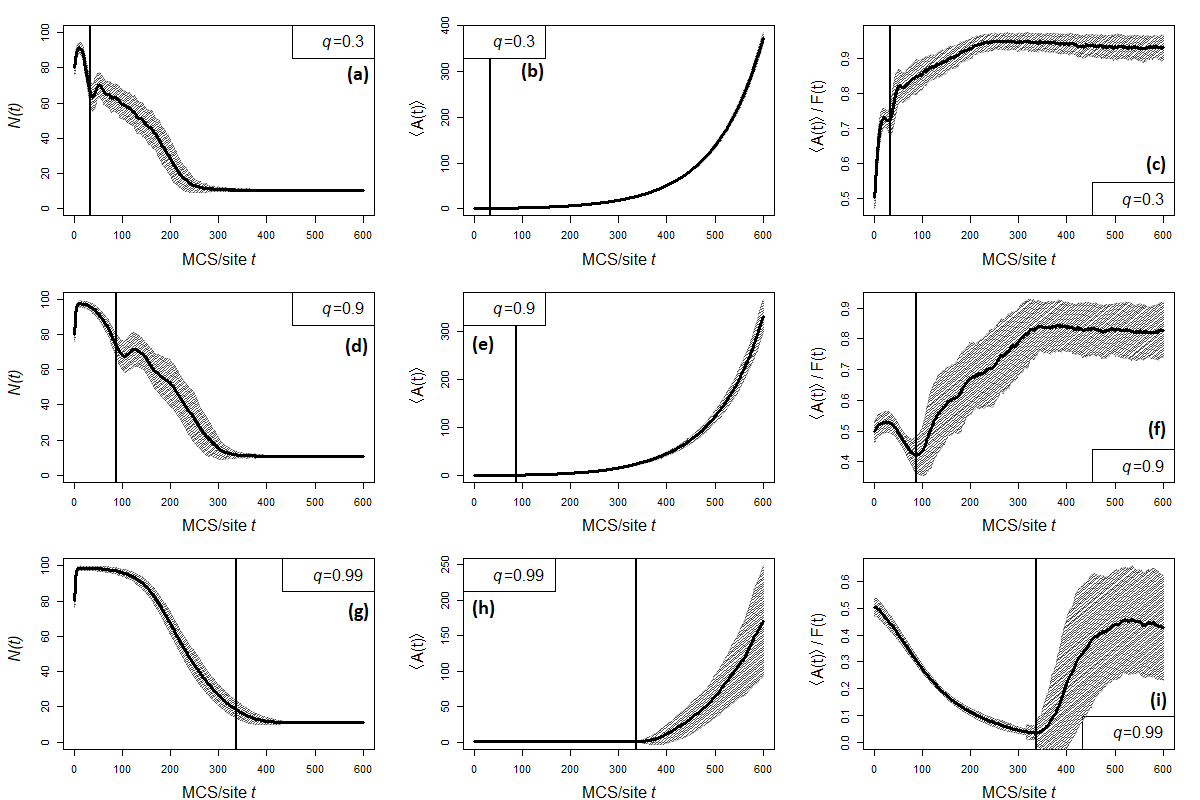}
\caption{Selected results for the egalitarian intervention case. Three columns show the values of the number of firms $N(t)$ (panels (a),(d),(g) in the first column), the average technology $\langle A(t) \rangle$ (panels (b),(e),(h) in the second column) and, the average to frontier technology ratio $\langle A(t) \rangle / F(t)$ (panels (c),(f),(i) in the third column) as functions of time. The results shown are for different values of the intervention probability $q$, with $q=0.3$ in the first row, $q=0.9$ in the second row, $q=0.99$ in the third row. The vertical straight line marks the time $t_c$ at which $\langle A(t)\rangle \geq F(t=0)=1$ for the first time. The average values over 400 simulations are shown with the black line and the $\pm 1$ SD over the simulations is represented as the grey area.}
\label{rys:9plotsR}
\end{figure}

Figure \ref{rys:9plotsR} presents some of the results for the egalitarian intervention case. 
First of all, as the probability of government intervention $q$ grows, the first stage lasts longer. In the egalitarian case, any company can be a beneficiary; in particular, a company with a low technology level. Therefore it takes more time for those companies to disappear from the system. This makes the initial growth of the average technology level slower. This can be seen in Fig. \ref{rys:9plotsR} when comparing the shape of $\langle A(t) \rangle / F(t)$ in the first 100 steps for different values of $q$. For sufficiently high $q$, the average to frontier technology ratio is dipping instead of growing since the frontier technology $F(t)$ grows much faster (exponentially) than $\langle A(t) \rangle$.

The most surprising results are for the very high intervention probability  $q=0.99$. From Fig. \ref{rys:9plotsR} we see that when the system enters the regime with $\langle A(t) \rangle \geq 1$ as in formula \ref{eq:p}, the technology growth becomes very unstable as indicated by the high standard deviation. At the time $t=600$, with higher $q$, the market has, paradoxically, a lower average technology compared to the situations with lower intervention probabilities $q$.

\subsection{High technology intervention}\label{section:high}

For the case in which the government supports only the firms with technology over one standard deviation ($SD$) above the average, there are no visible differences from the $q=0$ case shown in Fig. \ref{rys:baza}. This type of intervention does not influence the market significantly because the firms with high technology are already not likely to go bankrupt as follows from Eq. (\ref{eq:p}). Hence, there is a negligible number of government interventions involved in this case. For this reason, we have not visualized the results in the form of plots.

\subsection{Low technology intervention}\label{section:low}

\begin{figure}
\centering
\advance\leftskip-0cm
\includegraphics[width=153mm, height=130mm]{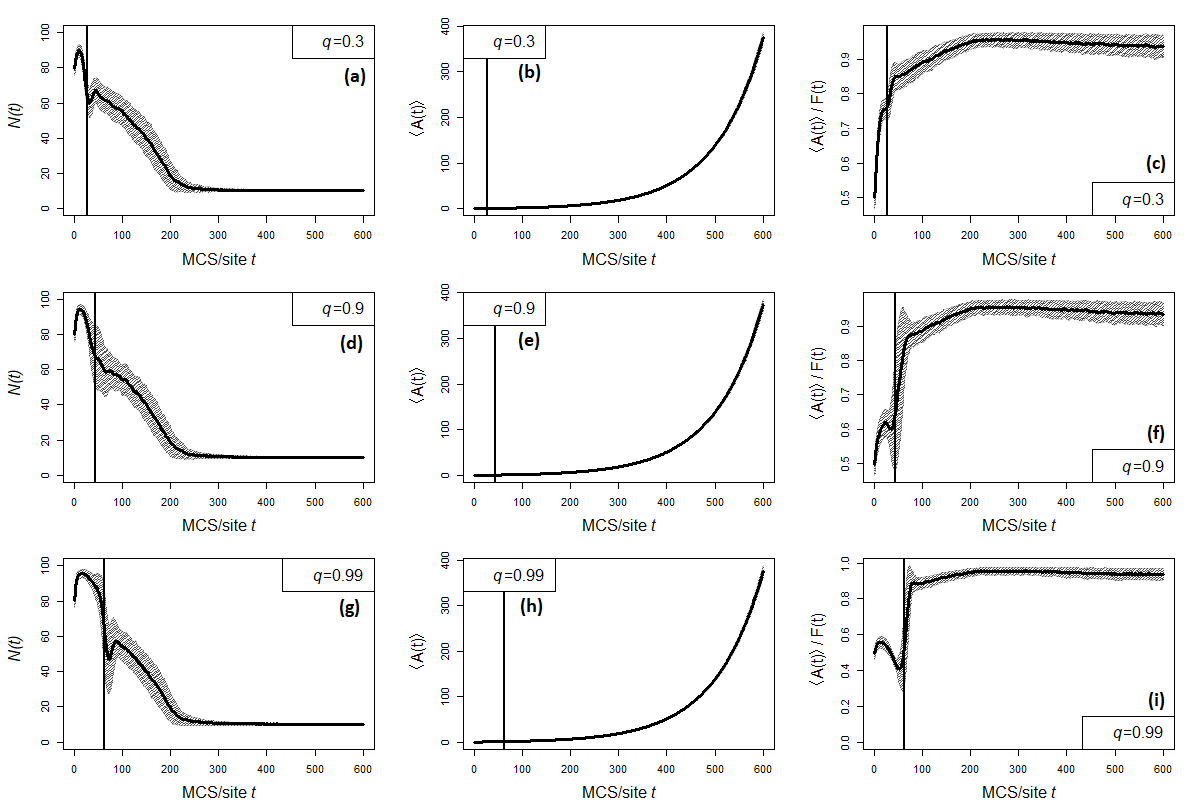}
\caption{Selected results for the low technology intervention case. Three columns show: (i) the values of the number of firms $N(t)$ (panels (a),(d),(g) in the first column), (ii) the average technology $\langle A(t) \rangle$ (panels (b),(e),(h) in the second column), and (iii) the average relative to frontier technology ratio $\langle A(t) \rangle /F(t)$ (panels (c),(f),(i) in the third column) as a function of time. The results shown are for different values of the intervention probability $q$, with $q=0.3$ in the first row, $q=0.9$ in the second row, and $q=0.99$ in the third row. The vertical straight line marks the time $t_c$ at which $\langle A(t)\rangle \geq F(t=0)=1$ for the first time. We show the average values over 400 simulations with the black line together with the band $\pm 1~SD$ marked by the grey color.}
\label{rys:9plotsL}
\end{figure}
When the government supports only the firms with low technology level, then the first stage, when typically low technology firms are usually eliminated, is prolonged. It is similar to the egalitarian intervention case. However, in contrast to the egalitarian intervention case, once those protected firms are eliminated, the growth continues undisturbed. That is, the ratio $\langle A(t)\rangle /F(t)$ is almost unaffected for the third stage.

Moreover, Fig. \ref{rys:9plotsL} shows that for about 100 steps, the number of firms in the system and the average technology to frontier ratio oscillate. After the weaker firms disappear, the dynamic is very similar to that shown in Fig. \ref{rys:baza}.

\subsection{Medium technology intervention}\label{section:medium}

\begin{figure}
\centering
\advance\leftskip-0cm
\includegraphics[width=153mm, height=130mm]{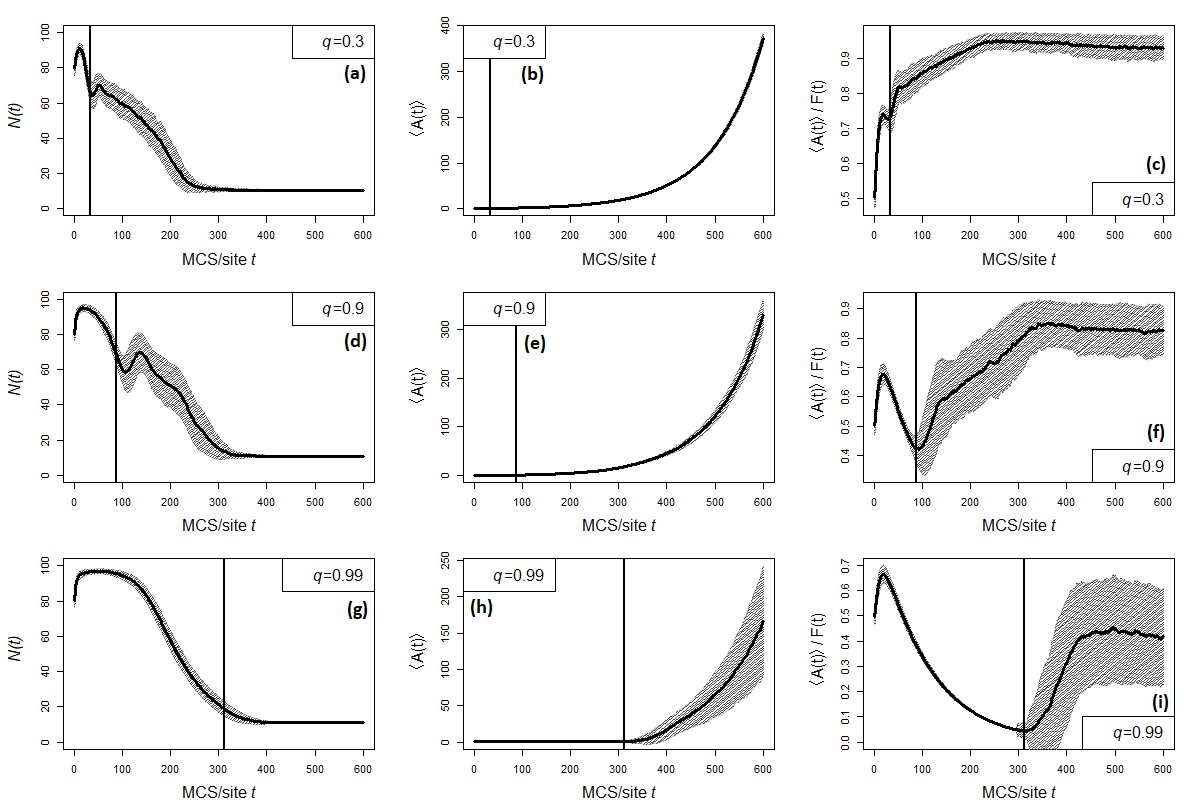}
\caption{Selected results for the medium technology intervention case. Three columns show the values of the number of firms $N$ (panels (a),(d),(g) in the first column), the average technology $\langle A(t) \rangle$ (panels (b),(e),(h) in the second column) and the average to Frontier technology ratio $\langle A(t) \rangle / F(t)$ (panels (c),(f),(i) in the third column) as functions of time. The results shown are for different values of the intervention probability $q$, with $q=0.3$ in the first row, $q=0.9$ in the second row, $q=0.99$ in the third row. The vertical straight line marks the time $t_c$ at which $\langle A(t)\rangle \geq F(t=0)=1$ for the first time. The average values over 400 simulations are shown with the black line and the $\pm 1~SD$ over the simulations is represented as the grey area.}
\label{rys:9plotsM}
\end{figure}
The government's aid towards the firms with the medium technology level, namely between $\pm 1~SD$, can have a visible impact on the growth dynamics of the market. Figure \ref{rys:9plotsM} shows that high intervention probability, in this case, destabilizes the market growth in the later stage. This is due to huge fluctuations expressed by the very high $SD$ value. The first stage lasting about 50 steps is undisturbed and is the same as the no intervention case in Figure \ref{rys:baza}. In this stage, mostly the weaker firms are going bankrupt as the government rarely intervenes. In the second and third phases, the dynamic is similar to the one with random intervention shown in Figure \ref{rys:9plotsR}. This should be expected since the low technology firms have already disappeared, and the intervention in the high segment does not influence the dynamic. Therefore, it is the intervention in the medium segment that can potentially have the most significant impact on the growth dynamic in the long term.


\subsection{The unstable growth for the strong intervention case}

Presumably, one of the most interesting results is a rapid increase of the quantity $\langle A(t) \rangle /F(t)$ vs. time $t$ (in MCS/site) for $q=0.99$, as shown in Figs. \ref{rys:9plotsR} and \ref{rys:9plotsM}. After the system reaches the range $\langle A(t)\rangle \geq 1$, the formula for the survival probability changes according to Eq. (\ref{eq:p}) and the system enters a new (last) stage. That is, the survival probability goes from the definition given by the first and third lines in Eq. (\ref{eq:p}) to the second and fourth lines. Hence, the average technology grows rapidly in an unstable manner in the $q=0.99$ case, as its large fluctuations are observed herein (i.e., the large standard deviation $SD$). This sudden change to a highly fluctuating dynamic of $ \langle A(t) \rangle /F(t)$ requires an explanation, which we provide below. However, the question why the dynamic switches to a rapid growth only after the technology $\langle A(t)\rangle \geq 1$ is still a challenge.

The origin of the behavior mentioned above lies in the way the intervention was implemented in the algorithm (see Sec. \ref{section:algor} for details) imitating a real interaction. This results in two effects that we list below. 
\begin{itemize}
\item[(i)] After the intervention saves a firm (see item 3 for details), the algorithm goes straight to item 7. It means that the firm behaves passively. The saved company can not do anything else at this Monte Carlo step.  This passive behavior happens very often for high $q$, since the intervention is very likely at any time. If the concentration of companies is greater than the minimum concentration (equals $\frac{N_ {min}} {L_x\cdot L_y})$, then instead of item 5 of the algorithm, the less effective item 6 of the algorithm works.  It results in companies' technology growing significantly slower than $F(t)$ or almost in technological stagnation. As a result the quantity $\langle A(t)\rangle /F(t)$ decreases, since $F(t)$ grows (exponentially) according to Eq. (\ref{eq:F}).
\item[(ii)]Near the $t=t_c$ threshold (when $\langle A_i(t)\rangle >1$ for the first time), the situation already changes because the concentration of companies is close to minimal (cf. Figs. \ref{rys:9plotsR} -- \ref{rys:9plotsM}, especially for plots with $q=0.99$). At this point, we are in the higher technology phase (i.e., $\langle A_i(t)\rangle >1$). Hence, the probability of company survival $p_i(t)$ is given, in general, by the third and fourth lines in Eq. (\ref{eq:p}). In principle, the inequality $A_i(t)> F(t)$ never fulfills, so only the third line works effectively. Since the concentration of companies is low, the external technology diffusion (defined in item 5 of the algorithm) becomes a possibility for the firms to increase their technology level. The meeting of two companies (located in neighboring lattice sites) is improbable, causing {\it de facto} disabling of item 6. The relative mean value $\langle A(t)\rangle /F(t)$ achieves the plateau. It happens when the increase of mean value $\langle A(t)\rangle $ is (to a good approximation) in a hundred percent caused by a global (external) frontier.
\end{itemize}

We can conclude that the dominant process for the case $\langle A(t_{MCS})\rangle > 1$ is described in item 5 of the algorithm (see Sec. 2.2 for details). It leads to a rapid increase in mean technology $\langle A(t_{MCS})\rangle $ and next to the plateau of its relative value.

\begin{figure}[H]
\centering
\advance\leftskip-0cm
\includegraphics[width=83mm, height=70mm]{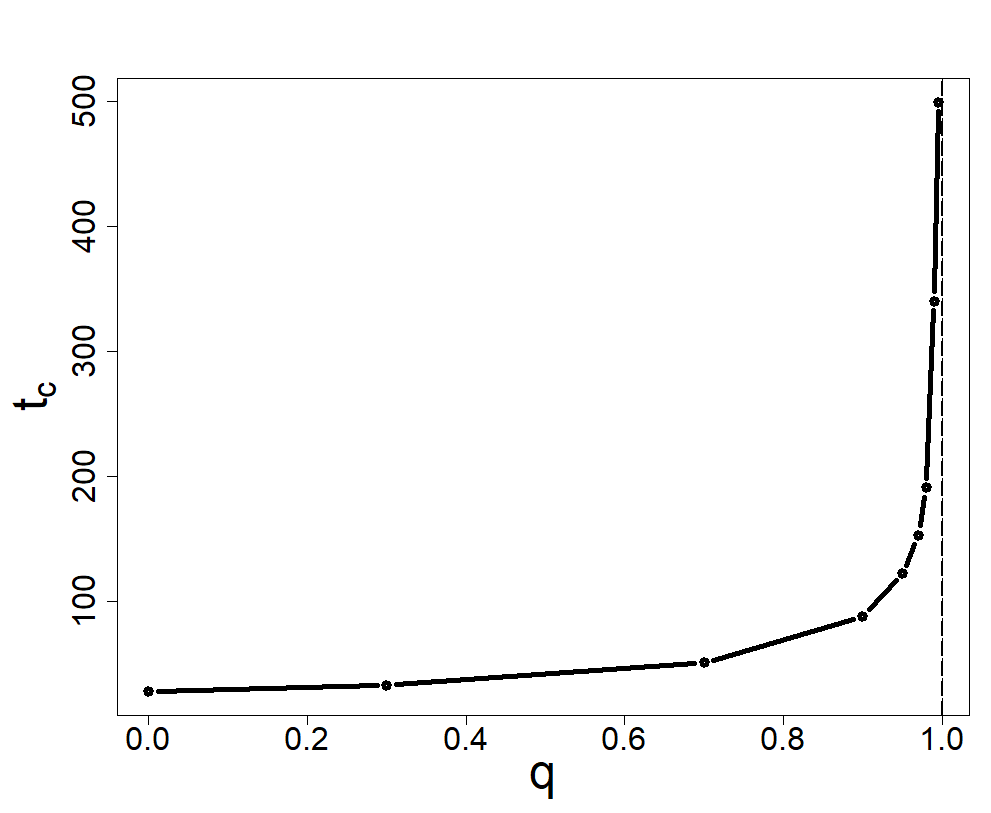}
\caption{Specific time $t_c$ at which $\langle A(t)\rangle \geq F(t=0)=1$ for the first time, as a function of the probability of the government intervention $q$.  The results were averaged over 400 simulations (or replicas) in this case. Despite the increasing help of the government parametrized by $q$, the time needed for the market to catch up to the initial world frontier technology level rapidly gets longer.}
\label{rys:tcsvsq}
\end{figure}

Moreover, we investigated the specific time $t_c$ for the unmodified algorithm at which the transition ($\langle A(t) \rangle \geq 1$) occurred and depicted it in Fig. \ref{rys:tcsvsq}. The result obtained is very characteristic because it shows that the level of market technology will never exceed the particular threshold $\langle A(t_{MCS})\rangle =1$ as the level of intervention tends to 100\%.

Figure \ref{rys:tcsvsq} shows that as $q$ grows, it takes the market more time to reach the high technology phase in which $\langle A(t) \rangle \geq 1$. When $q$ approaches 1, $t_c$ goes to infinity, i.e., the market will never reach the high technology phase. 

When $q=1$, the firms never bankrupt (cf. item 3). They can only disappear via merging with neighboring firms. Besides, new firms are created in the form of spin-offs (cf. item 6). 

\subsubsection{Remarks on the modified version of the algorithm}

\begin{figure}
\centering
\parbox{7cm}{
\includegraphics[height=7cm,width=7cm]{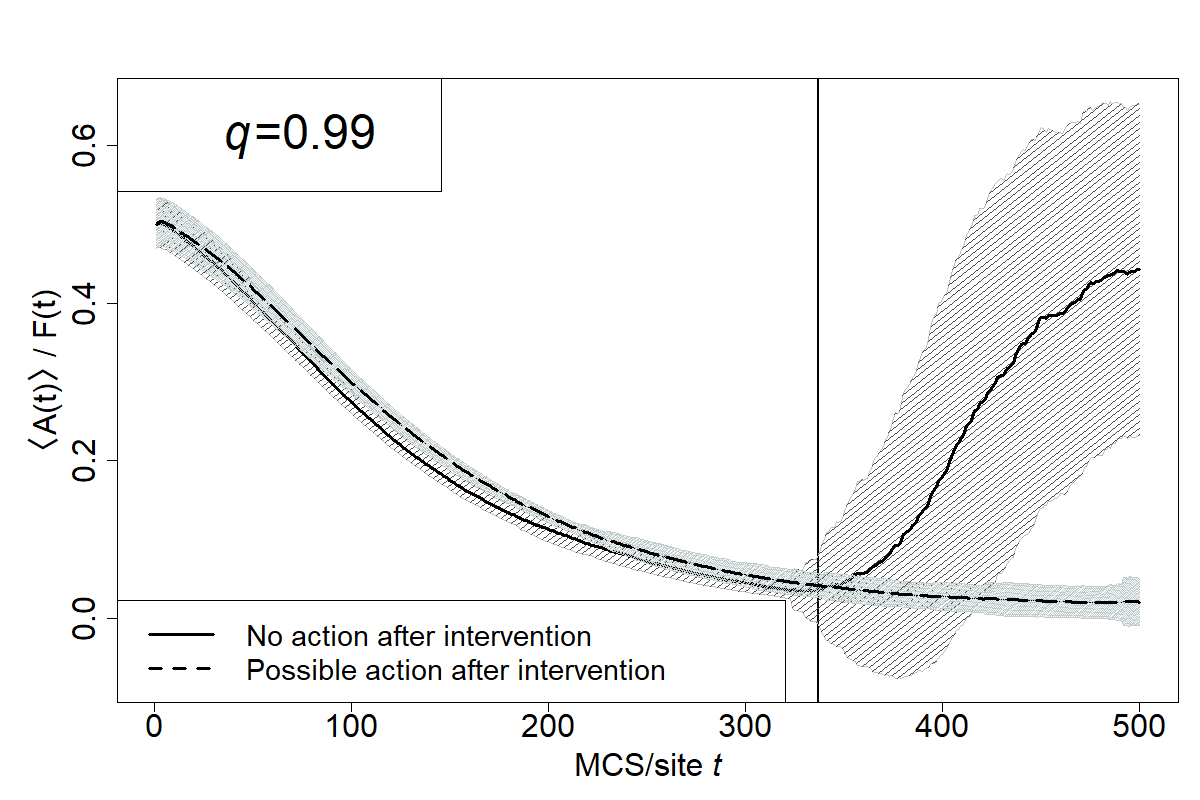}
\caption{Average technology to global frontier ratio for $q=0.99$. The solid curve is the result when the firms can not move after they were supported by the government. The dashed line concerns the case when firms can move and act after the intervention. Averaged over 400 simulations (or replicas) is sufficient herein. The vertical straight line marks the time $t_c$ at which $\langle A(t)\rangle \geq F(t=0)=1$ for the first time.}
\label{rys:altaf}}
\qquad
\begin{minipage}{7cm}
\includegraphics[height=6cm,width=7cm]{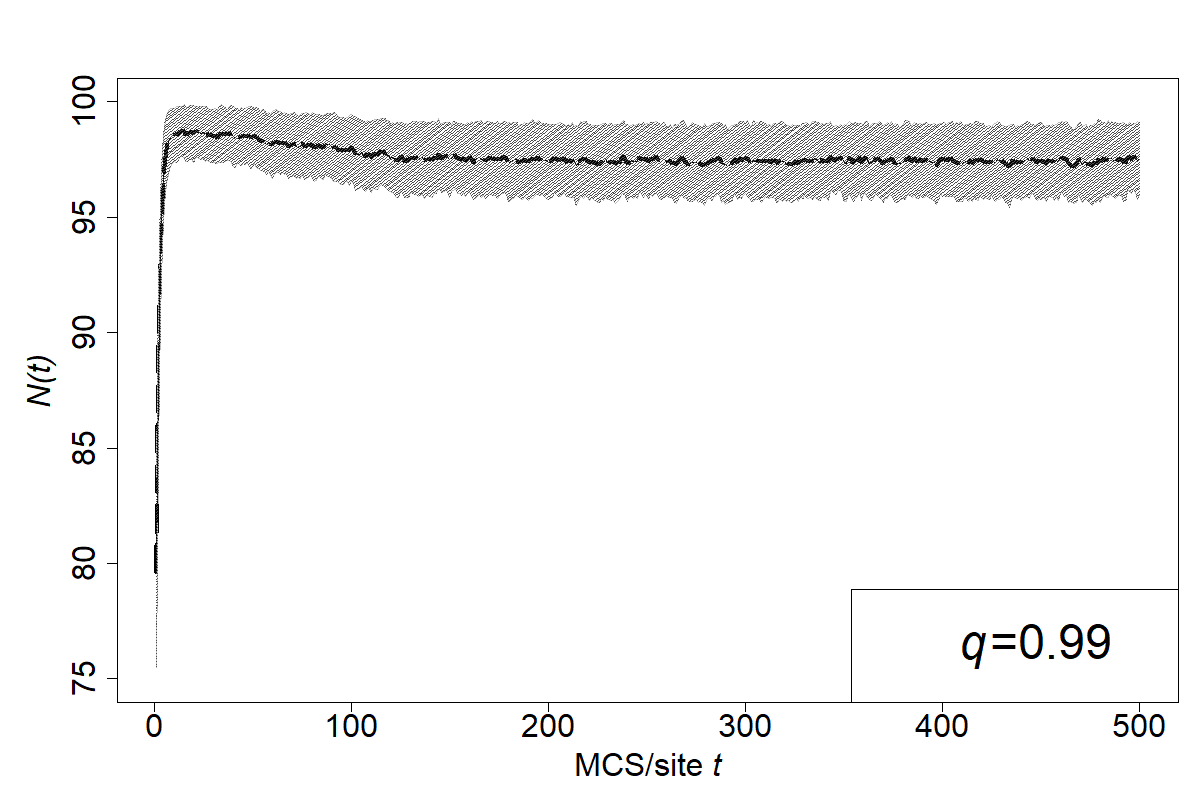}
\caption{Number of firms $N(t)$ for $q=0.99$. The case when the firms can move and act after the intervention. The results of this modified algorithm are averaged over 400 simulations (or replicas).}
\label{rys:Nalt}
\end{minipage}
\end{figure}

To further investigate the source of the unstable growth of the $\langle A(t)\rangle /F(t)$ value, we modified the item 3 of the algorithm. The modification is as follows: after the intervention, the firm can still act, which means the algorithm goes to item 5 instead of 7. Hence, the phase with unstable technology growth disappears\footnote{More precisely, $t_c$ is much larger (longer than 2000 MCS/site) in this case.}, which we show in Fig. \ref{rys:altaf} by the dashed curve. We emphasize that both algorithms can describe the real situation. It is because there may be companies that, after the intervention, "rest on their laurels" (unmodified algorithm) as well as others that invest in new technologies (the modified version of the algorithm).

The result obtained by the modified algorithm we present in Fig. \ref{rys:Nalt}. We show that the number of firms in the system is on a very high, stable level throughout the entire simulation. It is because new spin-offs are created more often than companies disappearing through merging or bankruptcy. Therefore, the firms in the market can rarely ever subject an external technology diffusion, which happens if they have no neighbors. The external diffusion is the way that allows for the fastest increase in technology in the long times. However, with large firm concentration, it is improbable. In the modified case, we observe that the system never enters a stage of fast $\langle A(t)\rangle /F(t)$ growth, which is shown by the dashed curve in Fig. \ref{rys:altaf}.



\section{Conclusions}

In our work, we considered two characteristic variants of the model of government interventionism in the companies' market. We describe interventionism by one parameter $0\leq q\leq 1$ only. It determines the level of interventionism. That is, the level of interventionism we define by the probability of government intervention $q$. The value of $q = 0$ indicates the absence of such interventionism, and the value of $q = 1$ indicates the certainty of interventionism. Notably, the government intervenes with probability $q$ if the company is threatened with bankruptcy, i.e., when the probability of bankruptcy is greater than zero.

Both model variants differ significantly in the behavior of companies following government intervention, although algorithms differ a little. In the first variant, after the intervention, the company is passive. In a given Monte Carlo step, its technological level remains unchanged. However, in the second (modified) variant, the firm is active, i.e., it can increase its technology level in the current Monte Carlo step.

In both model variants, only three different growth channels of the mean technology level of the companies' market exist. They are determined by items 5 and 6 of the algorithm (see Sec. \ref{section:algor} for details). These channels have a well-defined probability of appearing.

We show that government intervention leaves its mark on the growth dynamics of the company market. We found that for most of the values of $q$, government intervention does not influence the market significantly.  A significant impact of government interventionism takes place when $q$ is close to at least 0.9.

When the aid is given randomly, and the government does not distinguish firms based on their technology, it can have a disruptive effect on market growth. High intervention probability means the companies do not have to compete anymore because the government will almost always save companies from going bankrupt. The most drastic effects of the strong intervention are visible when one compares the value of the average to the frontier technology ratio $\langle A(t)\rangle /F(t)$ at $t=600$ for different $q$. With high $q$, $\langle A(t)\rangle /F(t)$ is volatile, and with asymptotic times its value is around 0.5, compared to the case with $q=0$ when it is close to 1. It means that the country with strong government intervention might never catch up with countries with lower intervention probability.

We also found that the volatile  $\langle A(t) \rangle / F(t)$ growth at long times disappears when the firms are allowed to act right after being saved by the government. 

We are aware that our model should also include the possibility of raising the technological level of companies through government interventionism. The positive effect of interventionism on technological expansion was observed in a study comparing Hong Kong and Singapore \cite{wang2018innovation}. However, at the moment, none of the studied variants of the model supply such a possibility.
We plan to investigate this potentially beneficial effect soon, as well as further extensions of the model. For example, we could add a budget variable to describe the firms more realistically and allow them to purchase assets from other companies to improve their technology and increase their market shares gradually.

\section{Acknowledgments}
We thank Professor Marcel Ausloos for an interesting discussion during the 10th Symposium FENS 2019 in NCBJ Otwock -- Swierk, Poland.

\bibliographystyle{plain}
\bibliography{bibliografiaartapp_final}
 
\end{document}